# $Co_2SeO_3Cl_2$: Studies of Emerging Magnetoelectric Coupling in a Polar, Buckled Honeycomb Material

Faith O. Adeyemi[1], Xudong Huai[1], Mohamed Kandil[2], Pradip Karki[2], Wencan Jin[2], and Thao T. Tran[1,3]*

[1]Department of Chemistry, Clemson University, Clemson, SC 29634, US

[2]Department of Physics, Auburn University, Auburn, AL 36849, US

[3]Department of Physics and Astronomy, Clemson, SC 29634, US

*thao@clemson.edu

**Abstract:**

The development of magnetoelectric materials requires chemical design strategies that integrate structural polarity with magnetic lattices capable of supporting competing spin interactions. Here, we demonstrate such an approach in the polar, buckled honeycomb magnet $Co_2SeO_3Cl_2$. Magnetization and heat-capacity measurements reveal strong magnetic anisotropy and four successive magnetic transitions at 25.4, 16.8, 11, and 3 K. The recovered magnetic entropy through the ordering regime is only around half of the expected $2R\ln(2)$, indicating persistent spin fluctuations. Second-harmonic generation measurements show three pronounced intensity anomalies at 11, 17, and 26 K that coincide with magnetic transitions while revealing that the crystallographic symmetry is preserved. Together, these results demonstrate that polar, buckled honeycomb magnets offer an unconventional phase space for coupling magnetic and electric dipoles in magnetoelectric materials.

## 1. Introduction

Frustrated magnets are of significant interest because competing exchange interactions at similar energy scales prevent spins from settling into a conventional magnetic order, potentially resulting in quantum fluctuations and novel states of matter.[1-16] Introducing electric polarity into such frustrated frameworks makes this class of materials even more compelling. A polar magnetic material displays an intrinsic, macroscopic electric dipole while promoting asymmetric magnetic exchange interactions in the presence of isotropic Heisenberg exchange.[17-20] Such polar, frustrated systems with appreciable asymmetric coupling can promote antiferromagnetic (AFM) and ferromagnetic (FM) interactions at comparable energies in ways that may be inaccessible in centrosymmetric counterparts. The ability to place magnetic spins in a polar, frustrated lattice opens a new path beyond conventional approaches to magnetoelectric coupling. Such materials can enable electric-field control of magnetic order, magnetic-field tunability of electric polarization, and enhanced sensitivity of both order parameters near phase transitions.[21, 22] However, achieving such magnetoelectric coupling remains elusive owing to an inherent energy-scale disparity between magnetic and electric dipoles. Recent studies on two-dimensional magnetoelectric materials, such as $NiI_2$ for electromagnon physics, light–spin coupling, and prospective opto-spintronic devices,[23-25] and $NiPS_3$ for optical spin readout and magneto-optical modulation, highlight the tunability accessible in reduced dimensions.[26-28] However, these systems often suffer from modest second-harmonic generation (SHG) signals. Three-dimensional materials, such as $Co_4Nb_2O_9$, $MnWO_4$, and $BiFeO_3$, provide greater SHG responses but typically exhibit less tunability than two-dimensional systems.[29-34] This trade-off underscores the need for a new chemical design strategy to realize polar, frustrated magnets that promote appreciable magnetoelectric coupling.

One effective strategy is to combine lone-pair-active building group with mixed-ligand coordination. Stereoactive lone-pair asymmetric subunits can induce global polarity if their local electric dipoles add to the overall polarization. At the same time, mixed ligands can generate local polarization at the magnetic site. To examine this design strategy, we create $Co_2SeO_3Cl_2$–a new polar, frustrated magnet that crystallizes in the monoclinic space group $P2_1$ and features a buckled honeycomb network of $Co^{2+}$ ions. The structure incorporates both $O^{2-}$ and $Cl^-$ ligands within the Co coordination environment, while the $Se^{4+}$ cation with its stereochemically active lone pair induces pronounced asymmetric distortions that enhance intrinsic polarity. Magnetic susceptibility measurements reveal pronounced magnetic anisotropy and four magnetic transitions. Heat capacity confirms the four transitions. SHG measurements confirm the non-centrosymmetric crystal structure and exhibit five phase transitions, of which three coincide with three magnetic transitions.

## 2. Results and discussion

### 2.1. Crystal Structure

$Co_2SeO_3Cl_2$ crystallizes in a non-centrosymmetric monoclinic lattice (space group $P2_1$). The magnetic $Co^{2+}$ atoms occupy two crystallographically distinct sites, forming buckled honeycomb layers, with the Co-Co distances ranging from 3.5127(6) to 3.9501(5) Å (Figure 1). The buckled honeycomb lattice is formed by alternating Co(1) and Co(2) within the *ac*-plane. Each Co atom is bonded to three O atoms of $SeO_3$ subunits and three Cl atoms, forming a distorted, polar octahedron [$CoO_3Cl_3$]. The Co-O bond distances range from 1.953(6) to 2.257(8) Å, and the Co-Cl bond lengths range from 2.450(4) to 2.506(3) Å. Each [Co(1)$O_3Cl_3$] unit is connected to [Co(2)$O_3Cl_3$] through edge-sharing along the *c*-axis and corner-sharing along the *a*-axis (Figure 1c). The $SeO_3^{2-}$ trigonal pyramidal group exhibits stereoactive lone-pair electrons with a local electric dipole. The Se atom has the +4 oxidation state with Se–O bond distances $1.711(7) < d < 1.720(49)$ Å. The local polarities of [Co(1)$O_3Cl_3$], [Co(2)$O_3Cl_3$], and $SeO_3^{2-}$ units add up, resulting in an overall polar crystal structure along the *b*-axis.

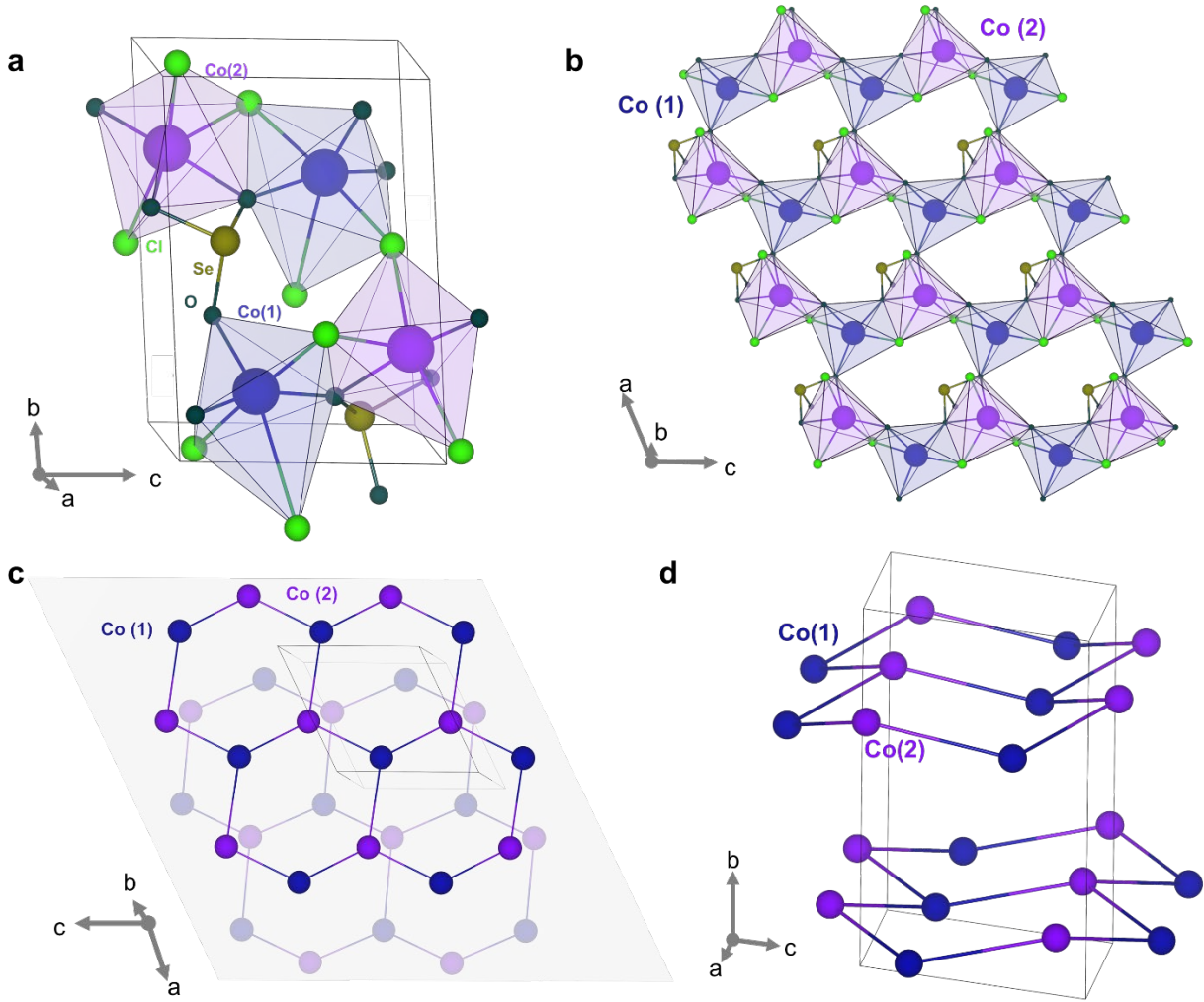

**Figure 1** (a) Crystal structure of $Co_2SeO_3Cl_2$; (b) Corner-sharing and edge-sharing network of $CoO_3Cl_3$ distorted, polar octahedra; (c) Honeycomb magnetic sublattice of the two distinct $Co^{2+}$; (d) Sideview showing the buckle honeycomb magnetic sublattice

## 2.2. Physical Properties

Temperature-dependent magnetization measurements (Figures 2a-b) along different crystallographic directions reveal pronounced magnetic anisotropy and four magnetic transitions at $T_{N1}$ = 25.4 K, $T_{N2}$ = 16.8 K, $T_{N3}$ = 11 K, and $T_{N4}$ = 3 K. Upturns in the $\chi(T)$ curves are observed after the onset of AFM ordering, suggesting competing AFM-FM interactions. Upon application of an external magnetic field, the transitions at $T_{N1}$ and $T_{N3}$ shift to higher temperatures, while those at $T_{N2}$ and $T_{N4}$ are suppressed to lower temperatures and eventually disappear by $\mu_0 H$ = 7 T. The shift to higher temperatures for the $T_{N1}$ and $T_{N3}$ transitions suggests that the spin entropy of FM coupling is reduced, which is compensated by an increase in the lattice entropy.

To account for the pronounced magnetic anisotropy arising from sizable spin-orbit coupling (SOC) of Co, we analyzed the magnetic susceptibility data using a modified Curi-Weiss (mCW) model.

[42-44] $\chi_\alpha(T) = \chi_{0,\alpha} + \frac{C_0}{T-\theta_\alpha} \frac{J_{eff}(J_{eff}+1)}{0.75} [w\, \mu_{1,\alpha}^2(T) + (1-w)\mu_{2,\alpha}^2(T)]$ (1)

$$C_0 = N_A \mu_B^2 / 3k_B \quad (2)$$

$$\mu_{i,\alpha}^2(T) = \frac{\mu_{0,i,\alpha}^2 + \mu_{1,i,\alpha}^2 e^{-\Delta_i/T}}{1 + e^{-\Delta_i/T}} \quad (i = 1, 2) \quad (3)$$

Where $\chi_\alpha(T)$ is the temperature- and orientation-dependent magnetic susceptibility, $\chi_{0,\alpha}$ is the temperature-independent diamagnetic and Van Vleck contribution, $\theta_\alpha$ is the modified Weiss constant for axis $\alpha$, $J_{eff}$ is the effective total spin number, $w$ is the fractional amplitude of Co(1) and Co(2), $\chi_{i,\alpha}(T)$ is the single-ion moment for Co on site $i$, and $\chi_{0,i,\alpha}$ and $\chi_{1,i,\alpha}$ are the orientation-projected moments of the ground state and the first excited state on site $i$, and $\Delta_i$ is the SOC gap.

The fit in the temperature range of 200 K $\leq T \leq$ 300 K yields characteristic single-ion energy scales of $\Delta_1$ = 148 K ($J_{eff}$ = ½ states populated) and $\Delta_2$ = 567 K ($J_{eff}$ = 3/2 states active and $J_{eff}$ = 5/2 unpopulated within the fitting temperature window). The resulting Weiss constants $\theta_a$ = -62(8) K, $\theta_b$ = -67(7) K, and $\theta_c$ = -28(1) K indicate a net AFM interaction and magnetic anisotropy (Table S1). The effective magnetic moment per Co extracted from the conventional Curie–Weiss analysis of a powder sample is 5.16 (7) $\mu_B$, which falls between the expected values 4.74 $\mu_B$ ($S$ = 3/2 and $L$ = 1) and 5.67 $\mu_B$ ($S$ = 3/2 and $L$ = 2). The orientation-dependent moments extracted are 5.6(1), 5.5(3), and 4.4(6) $\mu_B$, with $\mu_0 H$ // a, b, and c-axis, respectively. The largest value obtained along $a$ likely arises from enhanced orbital contributions, as evidenced by the extracted effective g-factor $g_{eff}$ = 2.9 along $a$, larger than $g_{eff}$ = 2.5 along $b$ and $g_{eff}$ = 2.2 along $c$. Overall, the $g_{eff}$ values are significantly higher than the expected spin-only value $g$ = 2, confirming appreciable SOC.

$M(H)$ curves at different temperatures reveal linear correlations, indicating AFM dominant interactions. Figure 2d shows $\Delta S_{mag}$ derived from the Maxwell relation. The transition at $T_{N1}$ corresponds to a negative entropy change, whereas those at $T_{N2}$ and $T_{N3}$ are associated with positive entropy changes. The transition at $T_{N4}$ first exhibits a negative entropy change at low fields, but then shows a positive entropy change at higher fields.

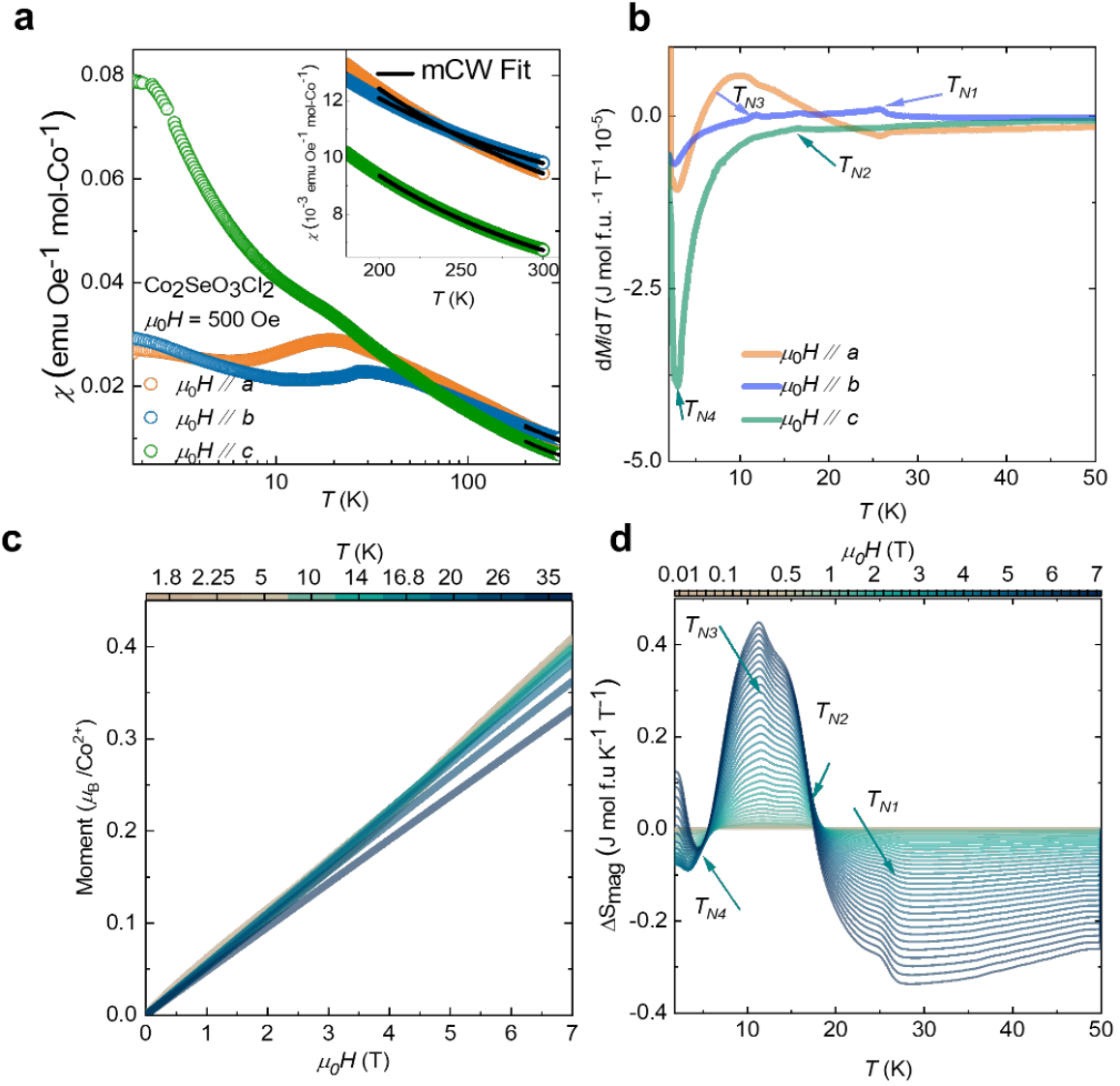

**Figure 2.** $Co_2SeO_3Cl_2$ (a) Orientation-dependent magnetic susceptibility vs. temperature with modified Curie-Weiss analysis at $\mu_0H$ =500 Oe; (b) Orientation-dependent d$M$/d$T$ vs. $T$ at $\mu_0H$ =500 Oe; (c) $M(H)$ of powder sample at different temperatures. (d) Magnetic entropy change ($\triangle S_{mag}$) at different magnetic fields at 1.8 K ≤ T ≤ 50 K

To confirm these observed magnetic transitions, heat capacity was measured on $Co_2SeO_3Cl_2$ with 1.8 K ≤ $T$ ≤ 300 K, and 0 T ≤ $\mu_0H$ ≤ 9 T. The heat capacity data confirm the four magnetic transitions observed from the magnetization data. At zero field, the phonon contribution was modeled by fitting the high-temperature heat-capacity data (45 K ≤ $T$ ≤ 300 K) using a sum of one Debye mode and two Einstein modes as follows:

$$\frac{C_p}{T} = \frac{C_{Debye(1)}}{T} + \frac{C_{Einstein(1)}}{T} + \frac{C_{Einstein(2)}}{T} \quad (5)$$

$$C_{Debye} = 9NRs_D \left(\frac{T}{\theta_D}\right)^3 D(\theta_D/T) \quad (6)$$

$$C_{Einstein} = 3NRs_E \frac{(\theta_E/T)^2 \exp(\theta_E/T)}{[\exp(\theta_E/T)-1]^2} \qquad (7)$$

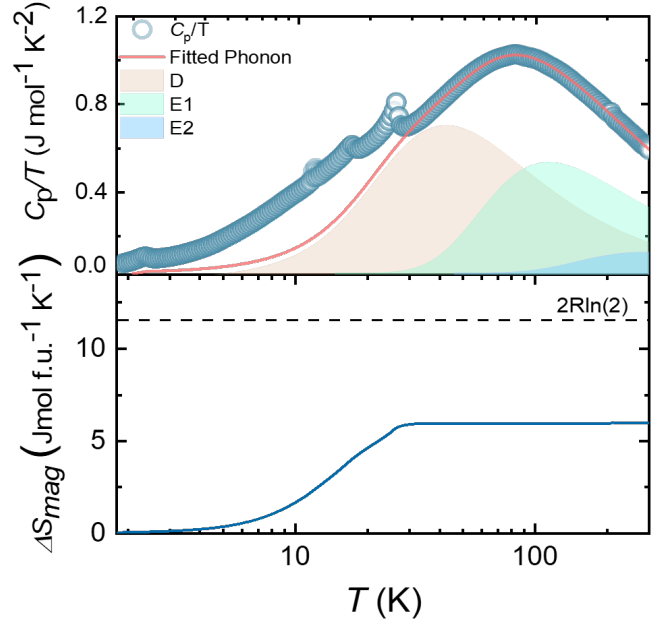

**Figure 3.** Molar heat capacity over temperature ($C_p/T$) vs. temperature for $Co_2SeO_3Cl_2$ at $\mu_0H = 0$ T and calculated phonon. The anomalies are consistent with the four magnetic phase transitions. Magnetic entropy change ($\Delta S_{mag}$) compared to the expected value of $J_{eff} = 1/2$ ($2*R\ln2$).

where $N$ represents the number of atoms, $R$ is the gas constant, $T$ is the sample temperature, $s_D$ is the number of oscillators of acoustic phonons, $\theta_D$ is the Debye temperature, $D(\theta_D/T)$ is the Debye function, $s_E$ is the number of oscillators of optical phonons. The fitted parameters are summarized in Table S1. The assessment of the phonon is based on the resulting good fit and physical oscillator terms. The fitted total number of oscillators is 7.7(8), close to the total number of 8 atoms per formula unit for $Co_2SeO_3Cl_2$. The presence of an Einstein contribution is justified by the characteristic maximum in $C_p/T^3$ vs. $T$ (Fig. S3). After subtracting the phonon contribution, the magnetic entropy change was estimated to be 5.9(2) J mol$^{-1}$ f.u. K$^{-1}$, only 51.2% of the expected value $2*R\ln(2) = 11.57$ J mol$^{-1}$ f.u. K$^{-1}$ and 25.6% of $2*R\ln(4) = 23.05$ J mol f.u.$^{-1}$ K$^{-1}$. The significant missing entropy observed from the analysis may arise from underestimated phonons or hidden entanglement between spins.

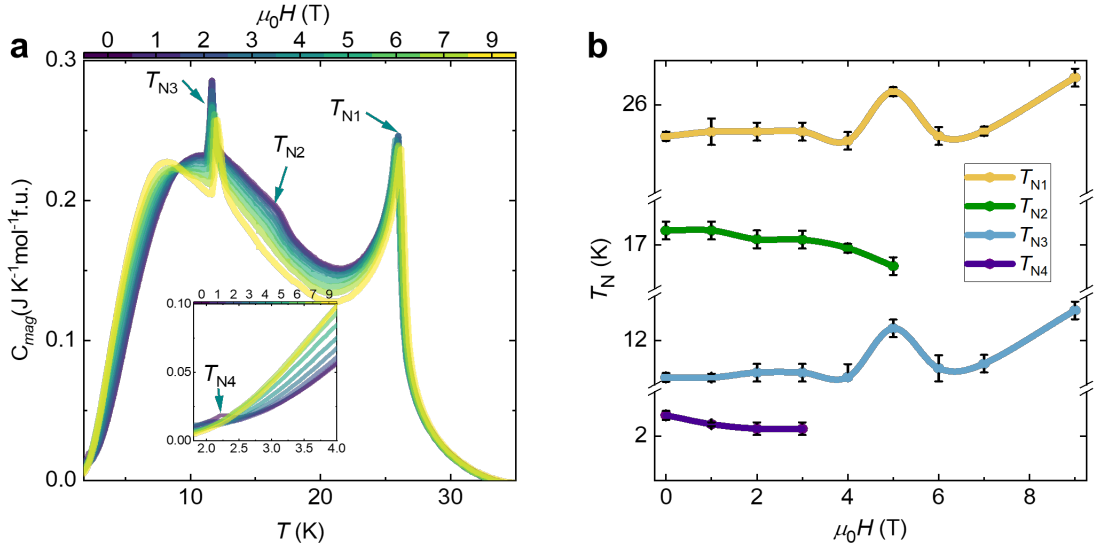

**Figure 4.** (a) Magnetic contribution to the heat capacity, $C_{\text{mag}}(T)$, measured under applied magnetic fields $\mu_0 H$ = 0–9T. The inset highlights the low-temperature region. (b) Field-dependence of the four transition temperatures.

The magnetic contribution to the heat capacity, $C_{\text{mag}}$, exhibits a pronounced field dependence. Figure. 4a shows how $C_{\text{mag}}$ evolves under applied fields between 1.8 and 30 K. Overall, transitions at $T_{N2}$ and $T_{N4}$ are progressively suppressed and eventually vanish, while $T_{N1}$ and $T_{N3}$ shift to higher temperatures as magnetic field increases (Figure 4b). This result is consistent with the magnetic susceptibility data.

## 2.3. Second Harmonic Generation

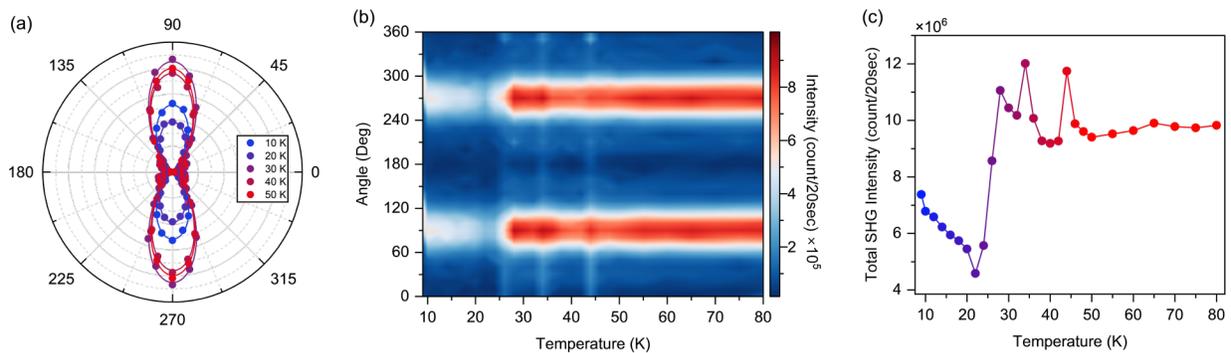

**Figure 5** (a) RA-SHG patterns measured in the parallel polarization configuration at selected temperatures across the magnetic transitions; (b) Two-dimensional colormap of the SHG intensity as a function of incident polarization angle and temperature; (c) Temperature dependence of the total SHG intensity obtained by integrating the signal over all polarization angles.

Rotational-anisotropy second-harmonic generation (RA-SHG) patterns were collected between 8 and 300 K, and the results are reproducible across consecutive thermal cycles. As shown in Figure

5a, the patterns can be fitted to a bulk electric dipole radiation in the polar $C_2$ point group. While the angular dependence of the RA-SHG patterns remains unchanged through the entire temperature range, anomalies in the SHG intensity are observed at multiple temperatures. As shown in Figures 5b-c, the most prominent feature is a sudden jump in SHG intensity at ~20 K during warming. In addition, multiple intensity spikes are observed at approximately 26 K, 34 K, and 44 K. These results suggest the presence of multiple phase transitions in the 10–50 K temperature window, which quantitatively modify nonlinear susceptibility tensor elements but do not change crystalline point group symmetry. It is noteworthy that the intensity spikes at 11, 20, and 26 K are in good agreement with the magnetic-driven AFM transitions, suggesting potential coupling between the electronic and magnetic dipoles.

## 2.4. Computational Studies

### Spin-Polarized Band Structure and Density of States

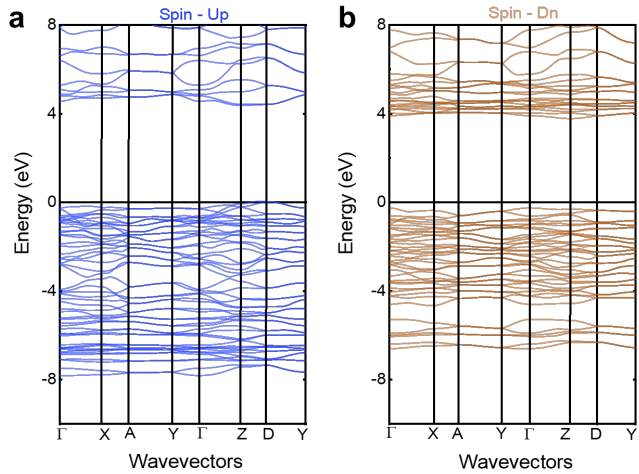

**Figure 6**. Spin-polarized band structure of $Co_2SeO_3Cl_2$

The electronic structure of $Co_2SeO_3Cl_2$ was investigated using spin-polarized DFT calculations. Strong hybridization between Co-d and O-p/Cl-p interacting atomic wavefunctions produces diffuse bands near the Fermi level ($E_F$), The density of states (DOS) shows that the majority spin (spin-up) of Co-d is fully occupied, while the minority spin (spin-down) is partially filled at the valence-band edge, consistent with the spin configuration of $Co^{2+}$ ($3d^7$, $S = 3/2$) (Figure 7). The valence band maximum is mostly dominated by Co-d, O-p, and Cl-p states. The O-p states are more dispersed compared to the Cl-p states because of the mixing between the O-p and Se-p states. The presence of Se-p states near $E_F$ indicates Se-s/p mixing of stereoactive lone-pair electrons. The conduction band minimum is mainly derived from the Co-d, O-p, and Se-p states.

The appreciable overlap between Co-d and O-p/Cl-p, along with the contrast in the bandwidths of O-p and Cl-p, may promote sizable competing exchange interactions, magnetic anisotropy, and magnetoelectric coupling.

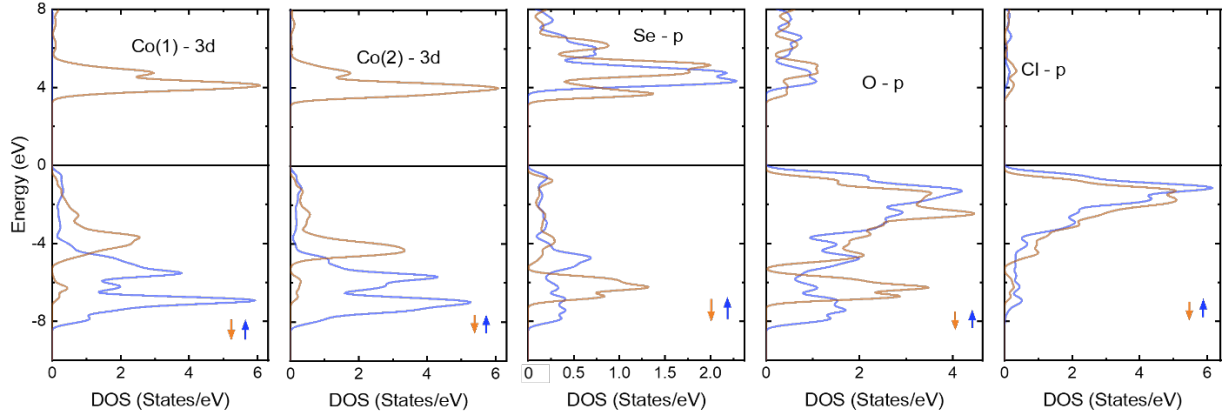

**Figure 7.** Spin-polarized density of states (DOS) of $Co_2SeO_3Cl_2$

## Chemical Bonding

Crystal orbital Hamilton population (COHP) analysis was performed on the relevant atom pairs (Co-O, Co-Cl and Se-O) to understand the relationship between bonding and physical properties in $Co_2SeO_3Cl_2$ (Figure 8).[37, 38] Co–O, Co–Cl, and Se–O bonds exhibit antibonding contributions (-pCOHP < 0) near the $E_F$. The electronic instability may prompt the system to evolve into new states of matter under external perturbations. To further quantify the bonding strength, integrated COHP (ICOHP) values were evaluated. Co–O bonds exhibit a larger (more negative) ICOHP value than the Co–Cl bonds, implying stronger hybridization. Se–O bonds display the largest (most negative) ICOHP value, indicating the most covalent bonding character compared to Co–O and Co–Cl bonds. Complementary COBI analysis reveals that Co–Cl and Se–O bonds exhibit almost a single bond (~0.9) and slightly more one-and-a-half bond (~1.6), respectively. These bond orders are higher than that of Co–O (~0.15). The ICOBI results reveal stronger covalent bonding character for Co–Cl and Se–O, and stronger ionic bonding character for Co–O. Overall, the ability to modify bonding characters within the polar, buckled honeycomb magnet using mixed ligands offers a chemical tool to promote cross-coupling between magnetic and electric components.

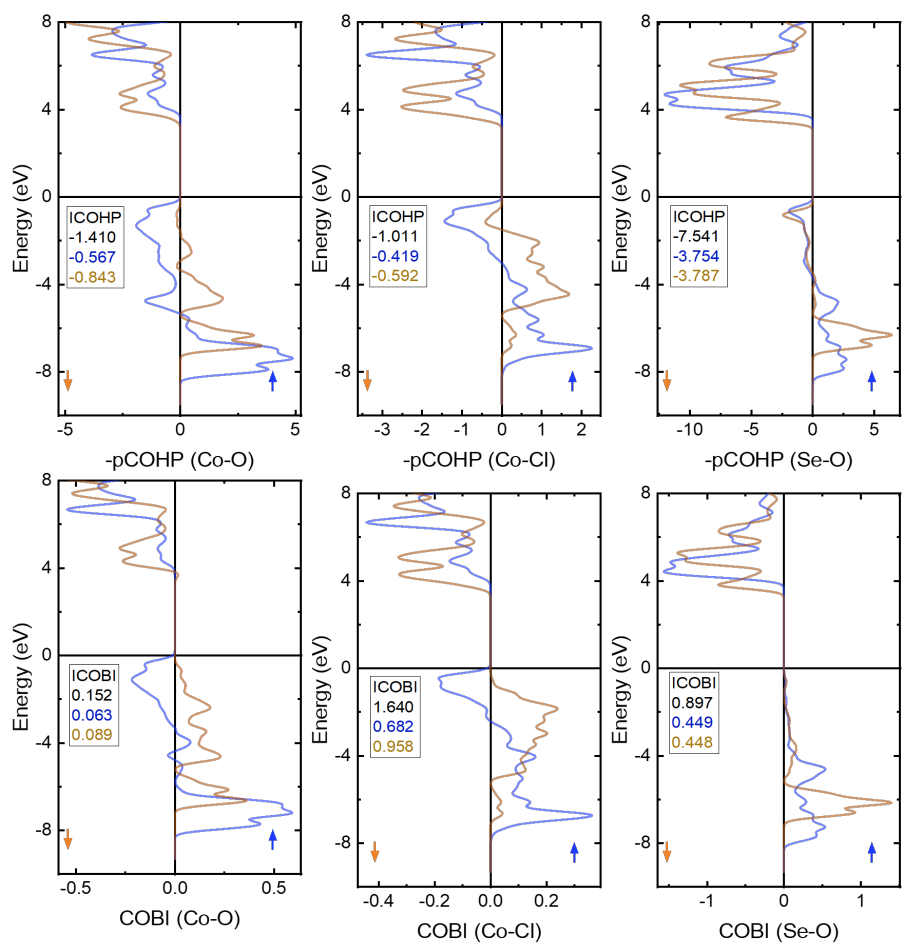

**Figure 8.** Crystal orbital Hamilton population (COHP) and integration (ICOHP), and crystal orbital bond index (COBI) for Co-O, Co-Cl, Se-O bonds in $Co_2SeO_3Cl_2$.

**Spin Density Map**

The spin density map projected onto the (010) plane shows that the spin is strongly polarized on Co-3$d$ and then polarizes O-p/Cl-p (Figure 9). The projection onto the (010) plane that cuts through a honeycomb layer reveals strong spin polarization, demonstrating the dominance of intralayered Co–O–Co and Co–Cl–Co superexchange pathways. On the other hand, the projection onto the (010) plane that goes between the honeycomb layers shows weaker but non-zero spin polarization, suggesting finite interlayered superexchange interactions.

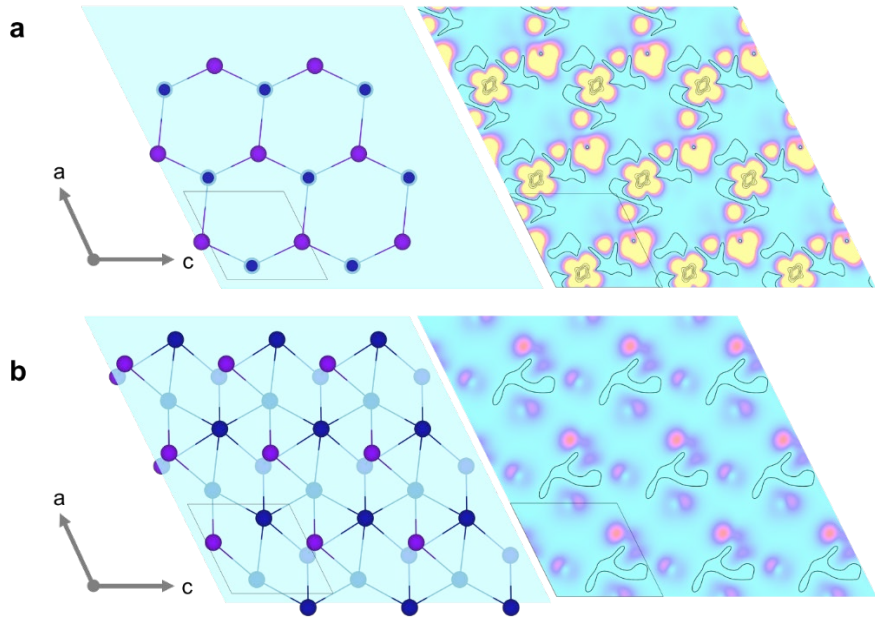

**Figure 9.** Spin density map of $Co_2SeO_3Cl_2$ (a) Projection onto the (010) plane cutting through the intralayer of Co; (b) Projection onto the (010) plane cutting through the interlayer of Co.

**Magnetic Exchange Interactions**

The calculated magnetic exchange interactions provide insight into intralayer and interlayer interactions (Figure 10). Within the honeycomb layer where Co-Co separations ~3.4-3.9 Å, intralayer magnetism is dominated by antiferromagnetic interactions with $J_1 = -49$, $J_2 = -45$, and $J_4 = -23$ K, respectively. Interlayer exchange interactions are generally weaker AFM interactions with $J_3 = -27$, $J_5 = -6$, and $J_6 = -5$ K, respectively. The resulting $J$ values are consistent with the observed exchange pathways in the spin-density map. We acknowledge that this calculation, based solely on the Heisenberg model, would not be the most accurate way to describe exchange interactions in the polar, frustrated magnet, where Dzyaloshinskii–Moriya and other anisotropic interactions may be important. The results are intended to provide some hints pointing to overall dominant AFM interactions, which are consistent with the magnetic properties.

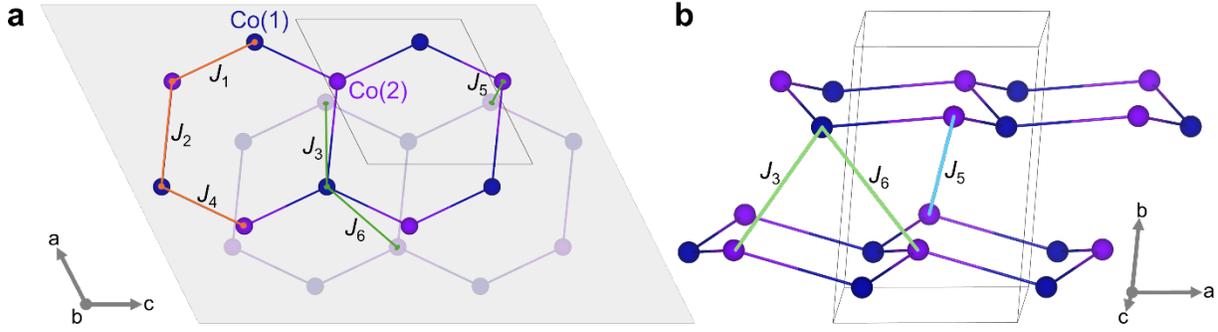

**Figure 10.** (a) Intralayer and (b) interlayer magnetic exchange interactions

## 3. Conclusion

Our results demonstrate that the polar, magnetic, buckled honeycomb material $Co_2SeO_3Cl_2$ offers an unconventional space for cross-coupling between magnetic and electric quantities. In contrast to the traditional planar honeycomb system, this material exhibits three common transitions, as observed in magnetic susceptibility and SHG measurements. These transitions are also confirmed by heat capacity data. In addition, $Co_2SeO_3Cl_2$ features mixed *J*-states and strong quantum fluctuations. Chemical bonding analysis shows the ability to modify bonding characters, bonding anisotropy, and magnetic and electric coupling pathways within this system. Additional characterizations are required to understand the microscopic origins of the potential emerging magnetoelectric coupling. This work establishes a new pathway to overcome the energy-scale mismatch and stabilize magnetoelectric interactions in ways that may be inaccessible in conventional systems.

**Acknowledgement.**

The work at Clemson University was supported by the NSF CAREER award NSF-DMR-2338014. X.H. and T.T.T acknowledge support from the Arnold and Mabel Beckman Foundation and the Camille Henry Dreyfus Foundation. W. J. acknowledges support by the Air Force Office of Scientific Research under Award No. FA9550-231-0499 and the NSF CAREER Award under Grant No. DMR-2339615.